# Overcoming contact resistance at metal–2D semiconductor interfaces: atomically clean $MoS_2$/Au ohmic junctions

Rafał Dunal[1]*, Maxime Le Ster[2], Iaroslav Lutsyk[1], Michał Piskorski[1], Paweł Dąbrowski[1], Paweł Krukowski[1], Witold Kozłowski[1], Aleksandra Nadolska[1], Wojciech Ryś[1], Paweł J. Kowalczyk[1] & Maciej Rogala[1]*

[1] Department of Solid State Physics, University of Lodz, Lodz, Poland

[2] HUN-REN Centre for Energy Research, Budapest, Hungary

* Correspondence to: rafal.dunal@edu.uni.lodz.pl, maciej.rogala@uni.lodz.pl

**Abstract**

The application of two-dimensional (2D) semiconductors, such as monolayer $MoS_2$, is limited by the high contact resistance commonly attributed to interfacial barriers at metal contacts. Furthermore, the dependence of electrical conductivity on $MoS_2$ thickness is still unsettled, as both increasing and decreasing trends with layer number have been reported. By showing the contrast between electrical transport of mono- and multilayer $MoS_2$ exfoliated on Au under ultra-high vacuum (UHV) and ambient conditions, we experimentally prove that, contrary to the prevailing view in the literature, the intrinsic $MoS_2$/Au junction is highly conductive and exhibits ohmic behaviour. Our results indicate that interfacial contamination is responsible for the high contact resistances reported to date and affects the thickness dependence of electrical transport, explaining the discrepancies observed in the literature. We rationalize those findings using electrical transport simulations. Lastly, we show that local force-mediated lamination on lightly contaminated contacts can recover pristine, ohmic contacts, offering a route towards nanoscale patterning.

**Introduction**

Atomically thin semiconductors, such as transition metal dichalcogenides (TMDs), have emerged as promising materials for next-generation electronics owing to their finite band gaps[1,2], strong electrostatic control[3], low-power operation and high on-off ratios[4]. These characteristics make them particularly attractive for post-silicon technologies, where continued scaling of conventional semiconductors becomes increasingly challenging[5]. At the same time, however, contact resistance is one of the major obstacles limiting the practical implementation of TMD-based electronics[6–8]. This issue is particularly critical for monolayer systems[9], which are the most attractive from the perspective of ultimate device scaling[10] but are frequently reported to exhibit poor electrical injection at metal interfaces[11–14]. In this work, we demonstrate that this limitation is not an intrinsic property of monolayer TMDs. Instead, we show that extremely low-resistance, nearly ohmic metal/TMD junctions can be achieved provided that the interface remains sufficiently clean during contact formation.

Among TMDs, $MoS_2$ is one of the most extensively studied and is commonly regarded as a representative material for this class of semiconductors[15,16]. It has been investigated for a broad range of applications, including sensors[17,18], light-emitting diodes[19], photovoltaics[20], resistive switching memory devices[21] and low-power field-effect transistors[22]. The intrinsic phonon-limited electron mobility in monolayer $MoS_2$ is predicted to reach several hundred $cm^2/V \cdot s$[23–25], and although this remains lower than in silicon[26], the combination of moderate mobility with excellent suppression of short-channel effects[27,28] makes monolayer $MoS_2$ highly attractive for aggressively scaled devices. In practice, however, the mobilities of monolayer $MoS_2$ devices are typically much lower, often not exceeding 50 $cm^2/V \cdot s$[29]. This discrepancy highlights that experimentally realised monolayer $MoS_2$ devices are strongly affected by extrinsic factors not fully captured in theoretical calculations, including stoichiometry, substrate interactions, material purity, and especially the quality of the metal/$MoS_2$ interface.

Numerous literature reports attribute the limited performance of $MoS_2$ devices to poor electrical contacts associated with the formation of substantial transport barriers at the metal/semiconductor interface[6–8,11,30]. Conductive Atomic Force Microscopy (CAFM) is well-suited for investigating such effects because it can map local vertical transport across mono-and multilayer regions on the same electrode. This approach minimises variations in interface quality between separately fabricated devices and therefore provides direct insight into thickness-dependent transport properties. Surprisingly, some of the CAFM studies reported that monolayer $MoS_2$ exhibits substantially higher resistance than thicker flakes[13], contrary to the intuition that the shortest out-of-plane transport path should facilitate carrier injection. At the same time, an opposite thickness dependence has also been reported, with thinner $MoS_2$ regions exhibiting higher conductivity than thicker ones[14], while still indicating barrier-limited transport at the interface. In addition, a pronounced increase in current with increasing CAFM probe load force is frequently observed[30–34] and is commonly interpreted as a consequence of strain-induced modifications of band alignment. All these observations imply that ultrathin $MoS_2$/metal interfaces inherently suffer from severe contact limitations, which may represent a major challenge for future electronic applications.

Here, we show that these issues are not universal. By forming the $MoS_2$/Au interface under pristine conditions, with Au cleaning, $MoS_2$ exfoliation and direct transfer all performed in UHV without exposure to ambient contamination, we obtain highly conductive, nearly ohmic monolayer behaviour. This contrasts sharply with the data commonly reported in the literature. In these samples, conductivity decreases with increasing $MoS_2$ thickness and shows no measurable dependence on CAFM probe load force.  To directly compare this behaviour with conventionally fabricated systems, we prepare an ex situ sample using standard tape exfoliation and polymer-assisted transfer in ambient atmosphere. This sample reproduces the behaviours typically reported in previous CAFM studies, including significantly high monolayer resistance, conductivity increasing with thickness, and pronounced pressure-induced current enhancement.

Using a resistance-network model based on Kirchhoff's laws, we qualitatively reproduce these opposite thickness dependences and show that they arise from differences in the $MoS_2$/Au interfacial resistance. The combined experimental and modelling results indicate that the high contact resistance commonly reported for monolayer $MoS_2$/metal systems does not primarily originate from intrinsic $MoS_2$/substrate interactions, but rather from contamination-induced barriers formed at the interface during sample preparation. This picture provides a unified explanation for previously reported pressure-dependent conductivity enhancement[30–34] and for the contradictory thickness-dependent transport trends observed in CAFM studies[13,14].

As a proof-of-concept demonstration of the role of interfacial contamination, we prepare an additional sample under an argon atmosphere, resulting in an interface of intermediate quality between pristine UHV and fully ex situ conditions. Although this junction initially exhibits a pronounced contact barrier, we demonstrate that nanoscale lamination induced by the pressure of the CAFM tip locally redistributes residual contaminants and permanently transforms the interface into a highly conductive state. Together, these results establish interfacial contamination as a decisive factor controlling whether monolayer TMD/metal junctions exhibit severe contact barriers or highly conductive, nearly ohmic behaviour. They therefore show that the contact-resistance bottleneck in two-dimensional electronics, widely reported in previous studies of ultrathin TMD/metal junctions, is not necessarily a fundamental materials limitation, but can be overcome by preserving or restoring the electronic quality of the metal/TMD interface.

## Results

The influence of interfacial contamination on the electrical properties of $MoS_2$ was investigated by comparing two types of samples, both consisting of $MoS_2$ flakes exfoliated from bulk crystals and transferred onto Au substrates.

### *Sample prepared in UHV conditions*

In the first case, the $MoS_2$/Au interface was prepared under strictly controlled UHV conditions to ensure an environment free from adsorbates that could contaminate either the electrode or the TMD surface. In this approach, the Au substrate was cleaned using repeated sputtering and annealing cycles in UHV, and subsequently, a freshly exfoliated $MoS_2$ crystal was brought into contact with the substrate within the same vacuum system (see Methods and Supplementary Fig. 1 for details). Performing the entire procedure without breaking the vacuum enables the formation of an interface that ensures an undisturbed contact between the two materials. This way, ultra-thin large-area $MoS_2$ flakes were obtained, appearing slightly darker than the Au substrate in optical images (Supplementary Fig. 2). Figure 1a shows the optical image of the structure selected for measurements, consisting of several monolayer terraces, where the area used for topographic and electrical characterisation is marked with a dashed line. The corresponding AFM image of the marked scan area is shown in Fig. 1b. AFM thickness measurements reveal domains containing 1 to 6 layers (Supplementary Fig. 3). Figure 1c presents a schematic representation of the scan area with individual heights expressed in layer units. This schematic is included to facilitate identification of regions with different thicknesses in the topography image (Fig. 1b) and the corresponding CAFM conductivity map acquired at a sample bias of +0.4 V (Fig. 1d).

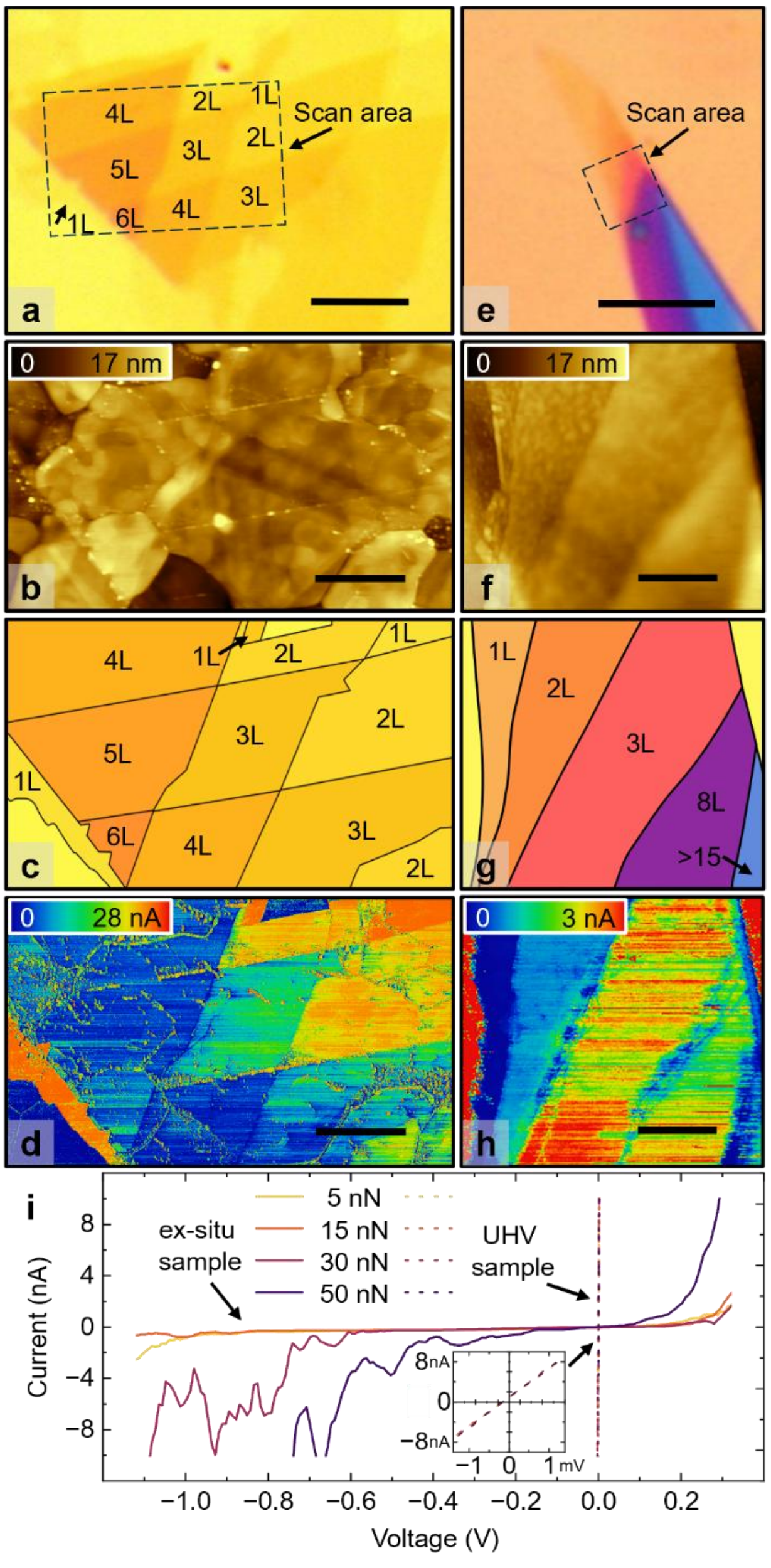

**Figure 1.** Comparison of UHV (a–d) and ex situ (e–h) samples. Panels (**a–d**) show optical image, semi-contact AFM topography map, height of particular regions, and CAFM current map for the UHV-prepared sample, while panels (**e–h**) present the corresponding results for the ex situ sample. CAFM current maps were acquired at 0.4 V (UHV) and 0.2 V (ex situ). Scale bars: 5 µm (**a**,**e**); 3 µm (**b**,**d**); 0.5 µm (**f**,**h**). (**i**) Representative current-voltage (IV) curves measured on monolayers for both samples with varying load force of the tip.

Following the sample fabrication under UHV conditions, it was subsequently moved to the AFM microscope in ambient air. During this exposure time, contaminants accumulated on the sample surface, requiring the scan area to be cleaned using CAFM scans[35,36]. These were sufficient to remove impurities on the $MoS_2$ regions but could not fully eliminate the contamination layer covering the exposed Au, which explains the poor conductivity visible in the bottom-left corner of Fig. 1d. Nevertheless, the interface quality was not affected by the contaminants, as they accumulated on the top of the structure and not between the Au substrate and $MoS_2$.

The highest conductivity is measured in the monolayer regions, depicted in Fig. 1d as red areas in the top-right and bottom-left corners (see also Fig. 1c). Subsequently, the measured current decreases as the thickness of $MoS_2$ increases up to 5 layers. While a slight decrease is observed for the bi- and trilayer, it decreases significantly for the fourth and fifth layer. Subsequently, the measured current increases for the six-layer region, which will be addressed later in this paper.

Such a finding is in contradiction with the previous report on layer-dependent conductivity in a similar $MoS_2$ system, which shows the opposite trend, with the highest energy barrier for the monolayer[13]. However, as shown in Fig. 1i, currents that we measured for the UHV monolayer are remarkably high considering the nanoscale tip–sample contact area, indicating a low $MoS_2$/Au contact resistance. Within the linear operating range of the CAFM system (up to ~10 nA), the current-voltage (IV) characteristics are strictly linear, evidencing ohmic behaviour of the junction. Their linearity is further corroborated by measurements over the full operating range, which show excellent quantitative agreement with IV characteristics measured for a reference resistor of comparable resistance to the value measured for the monolayer (Supplementary Fig. 4).

Previous reports also present point IV measurements as a function of tip load force, showing a significant increase in the measured current with increasing force[30–34]. In this case, however, no such effect was observed, and the IV characteristics were virtually independent of the force setpoint (Supplementary Fig. 5) and demonstrated excellent repeatability.

Our results align with the DFT calculations of Velicky et al.[37] and Blue et al.[38]. In both cases, monolayer $MoS_2$ on Au substrate is predicted to exhibit a metallic character resulting from an increase in the $MoS_2$ density of states at the Fermi level. It is explained by either electron transfer from Au to $MoS_2$[37] or hybridization of the S and Au orbitals[38]. Moreover, our results show that the Au-induced enhancement of the electrical transport properties persists also in the subsequent $MoS_2$ layers. This enhancement is particularly pronounced in the second layer, while it becomes noticeably weaker for the third layer (Fig. 1d and Supplementary Fig. 5). For four layers, the material clearly recovers its semiconducting character, as evidenced by distinctly nonlinear IV curves (Supplementary Fig. 5). These observations are corroborated by Raman spectroscopy measurements, which show evolution of modes up to the fourth layer, indicating an interaction with the substrate[39,40] (Supplementary Fig. 6).

***Sample prepared ex situ in standard conditions***

The same analyses were also carried out for a second sample prepared by micromechanical exfoliation and PDMS transfer onto Au in ambient conditions. In the following, this sample is referred to as the ex situ sample, and it represents the standard fabrication approach commonly used in the literature[41,42]. As such, it can be regarded as representative of the typical $MoS_2$/Au structures reported previously. It should be noted, however, that the exact degree of interfacial contamination can vary between studies, as it may depend on the exposure time of both the Au substrate[37] and the exfoliated $MoS_2$[43] to air before transfer, as well as on ambient humidity and the concentration of airborne hydrocarbons[44].

For topographical and electrical characterisation, a structure containing multiple terraces of different heights was selected, as presented in Fig. 1. Figure 1e shows the optical image of the $MoS_2$ flake and Fig. 1f is the topography map obtained by AFM. The AFM height measurements are summarized in Supplementary Fig. 7. Figure 1g shows a schematic representation of the scan area from Fig. 1e. Figure 1h presents a current map obtained by CAFM at a bias of +0.2 V.

These measurements show that conductivity increases with $MoS_2$ thickness, in stark contrast to the UHV-prepared sample and the mentioned DFT predictions of a metallic $MoS_2$ behaviour on the Au substrate[37,38]. In the ex situ sample, the conductivity measured for the first layer is extremely low, indicating the presence of a large barrier for charge transport at the $MoS_2$/Au junction. This observation is in line with previous reports in which monolayer $MoS_2$ exhibited an unexpectedly low conductivity[11,13,14,30–32], with some studies attributing this behaviour to the formation of a Schottky barrier at the $MoS_2$/Au interface[13,30,31]. The currents measured for the thicker layers are much larger (Fig. 1h), which is also in line with previously reported data[13]. This increase was initially attributed to a reduction in the Schottky barrier height associated with the increase in electron affinity with increasing $MoS_2$ thickness[13]. Moreover, the IV curves measured for an ex situ sample monolayer (in contrast to the UHV one) show an increase in current with increasing load force, as presented in Fig. 1i. A similar dependence has been reported in previous works[30–32] and was largely attributed to strain-induced changes in the band alignment in the $MoS_2$.

Since neither the suppressed monolayer current nor the load-force-dependent current increase is observed for the UHV-prepared sample, where the $MoS_2$/Au interface is clean, we attribute both effects in the ex situ sample to the presence of a thin insulating contamination layer at the interface[44]. In the monolayer regime, this air-borne interfacial layer can act as an additional tunnelling barrier that strongly limits charge transport, giving rise to the apparent large contact resistance, which was often interpreted in previous reports as resulting from a Schottky barrier[13,30,31]. Thus, the load-force-dependent current increase can be consistently explained by pressure-induced thinning of the contamination layer, which reduces its resistance at the measurement point. The presence of interfacial contamination can also explain the discrepancies in the current-thickness dependence between the two sample types, which will be addressed in the following section.

***Qualitative comparison of UHV- and ex situ prepared samples***

The measured resistance of $MoS_2$ layers shown in Fig. 2 for both sample types was extracted from CAFM current maps presented in Fig. 1 (see Supplementary Fig. 8 for details of the extraction procedure). For the UHV sample, the resistance increases with the number of layers (Fig. 2a), except for the six-layer case, which appears to be an anomaly. The ex situ sample on the contrary shows an opposite trend; the resistance decreases significantly for 2 layers and remains at a similar level for 3 to 8 layers (Fig. 2b). Such discrepancies between the two samples can be conceptually rationalized by qualitative simulations based on Kirchhoff's laws implemented within a 3D resistor network representing the resistances of multilayer $MoS_2$ structures (see Methods for details).

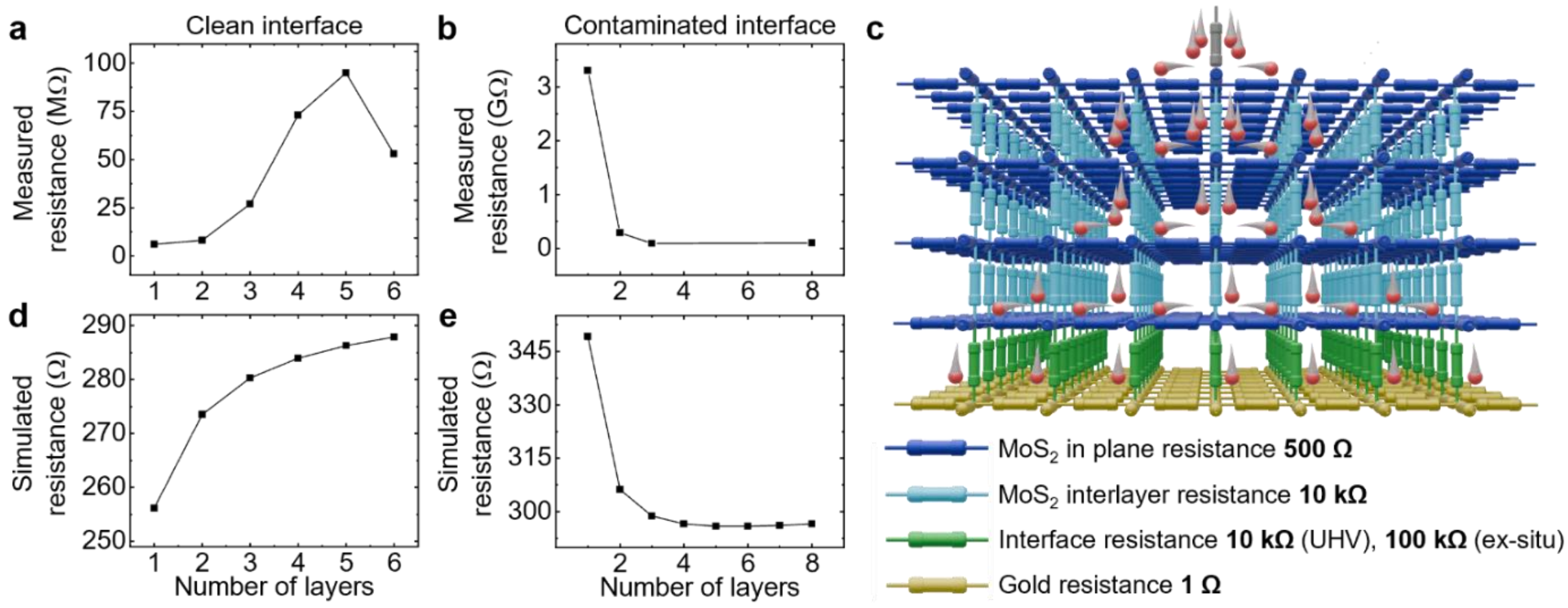

**Figure 2.** Resistance/layer number dependence for UHV and ex situ samples. (**a**) Measured resistance for UHV and (**b**) ex situ sample. The experimental resistance values were determined from the conductance analysis of the CAFM images shown in Fig. 1 (see Supplementary Fig. 7 for the analysis procedure details). (**c**) Resistance network simulation model with the values of resistance used for every resistor type. (**d**) Simulated equivalent resistance for low resistance interface system and (**e**) high resistance interface system.

In Fig. 2c, a schematic model of the resistor network used in our simulations is illustrated. The network consists of parallel resistor planes representing the in-plane resistance of the $MoS_2$ layers (dark blue), Au substrate (yellow), interconnected by out-of-plane resistors representing resistance between $MoS_2$ layers (cyan) and at the $MoS_2$ / Au interface (green). While such resistance network models are not meant to quantitatively simulate the charge transport physics of the $MoS_2$/Au junction, we show that they successfully explain the impact of the interfacial layer on the overall resistance trends in measured systems.

The simulation calculates the equivalent resistance of the resistor network based on the number of $MoS_2$ layers in the structure. Due to the high resistance of the $MoS_2$ layers, a connection of the top layer to the circuit in the central node simulates a point contact of the CAFM probe (symbolically shown in Fig. 2c as a grey resistor). In contrast, the highly conducting Au layer behaves as a globally connected electrode. Thus, current does not flow in a straight line between the tip and the Au but instead spreads over the $MoS_2$ layers, as illustrated in Fig. 2c by red balls symbolizing charge carriers. The overall current distribution in the network depends on the configuration of the nodes and the resistances between them, which define the equivalent resistance calculated in the simulation. It is therefore determined by the ratio between in-plane and out-of-plane resistances, as well as by the number of $MoS_2$ layers.

The resistance values used in the simulated resistor networks shown in Fig. 2c were chosen to reproduce the high conductivity anisotropy of $MoS_2$[45]. The two simulated 3D resistor networks, which are compared, differ only in the resistance of the $MoS_2$/Au interface (green) resistors. In the low Au/$MoS_2$ contact resistance system (Fig. 2d), the interface resistance was assumed to be equal to the interlayer resistance between $MoS_2$ layers, whereas in the high Au/$MoS_2$ contact resistance system (Fig. 2e), it was assumed to be ten times larger. The resistance values were chosen to reproduce the correct resistance proportions in the measured $MoS_2$/Au system; however, their absolute values should not be interpreted as real physical values.

In the low Au/$MoS_2$ contact resistance system (Fig. 2d), the equivalent resistance increases monotonically with each added layer, consistent with the experimental results for the UHV-prepared sample (Fig. 2a). In contrast, the high Au/$MoS_2$ contact resistance system (Fig. 2e) shows a decrease in equivalent resistance as the number of layers increases, in quantitative agreement with the ex situ sample electrical measurements (Fig. 2b).

In the case of the low Au/$MoS_2$ interface resistance network (Fig. 2d), the trend is explained by a contact resistance comparable to interlayer $MoS_2$ resistances. In this situation, the dominant contribution arises from the interlayer resistance, whose cumulative value increases with the number of $MoS_2$ layers. This leads to a gradual increase in the equivalent resistance of the entire system with increasing layer number, as shown in Fig. 2d, and explains the same trend observed for the UHV sample electrical measurements (Fig. 2a).

In contrast, for the high Au/$MoS_2$ contact resistance system (Fig. 2e), the main contributor of the overall resistance is the interfacial resistance. The explanation here is more complex and involves the number of Au/$MoS_2$ interface resistors forming parallel current paths. In the monolayer case, there is only one plane of in-plane $MoS_2$ resistors, so the lateral spreading of current is strongly limited. Consequently, the current will pass only through a limited number of interface resistors. As their resistance is high, the equivalent resistance stays high. When the number of simulated layers increases, the increased amount of in-plane resistors (Fig. 2c) provide additional low-resistance pathways, allowing the current to spread more effectively in the in-plane directions. As the charges spread, the current passes through a larger number of Au/$MoS_2$ interface resistors in parallel. As a consequence, the resistance of the entire interface layer decreases. This resistance drop outweighs the increase in resistance associated with the addition of successive out-of-plane resistor layers, leading to an overall decrease in the equivalent resistance. By analogy, this can explain the ex situ sample measurement results (Fig. 2b), where the increase in the number of layers is associated with an increase in the current-carrying cross section within the contamination layer. This leads charge carriers to access more conduction pathways in that layer, leading to a reduction in the overall resistance.

The qualitative simulations described above are sufficient to explain the differences in the layer-dependent conductivity of both samples; however, they do not account for all physical effects present in real systems. For the experimentally measured resistance of the UHV sample (Fig. 2a), a deviation from the behaviour predicted by the low Au/$MoS_2$ contact resistance model (Fig. 2d) can be observed. In the real system, the resistance of the first two layers remains at a similar level and increases significantly for subsequent layers. In contrast, in the simulated system, the increase in resistance between the first two layers is much more pronounced. We attribute this difference to additional physical effects that are not incorporated in our model based only on Kirchhoff's laws. It must be emphasized that the resistor network model does not account for the enhanced conductivity of the bottom $MoS_2$ layers present in the UHV sample, which may be related to the previously mentioned Au-induced increase in the $MoS_2$ density of states[37,38]. Additionally, an unexpected increase in conductivity is observed for six layers; however, this result should be interpreted with caution, as the mapped area is substantially smaller than for the other layers, limiting the statistical robustness of the conductivity data.

***Interface improvement by nanoscale lamination***

To directly confirm the influence of contaminants, we also demonstrate that the interface properties can be locally improved, enabling the transformation of a high-resistance junction formed under non-optimal conditions (outside UHV, with exposure to foreign molecules) into a high-conductivity junction preferred in electronic devices. Through local nanoscale lamination, interfacial impurities can be relocated, thereby modifying the $MoS_2$/Au interface. This impacts the electrical properties of the material and locally switches the interface from a high-resistivity state to a more conducting one. To achieve that, we used a sample for which $MoS_2$ was exfoliated and transferred onto Au using a standard process (as for the ex situ sample), but under controlled atmosphere conditions (in a glovebox) after preheating the Au substrate at 373 K to minimise residual moisture on the surface. This approach significantly reduces interfacial contamination and can be readily translated to scalable industrial processes. Next, we conducted a series of scans with varying tip load forces using the local force spectroscopy mode[46,47]. Although related strategies have been presented in earlier reports[48–52] they have so far been limited to implementations based on contact-mode AFM. The use of force-spectroscopy methods could be particularly beneficial for mechanically compliant materials, as it substantially reduces the contribution of lateral shear forces that are present during scanning in contact mode[53].

Results of our experiment are shown in Fig. 3a, which presents a sequence of images illustrating the progressive modification of the $MoS_2$ flake. In the initial state (for load forces ≤120 nN), the surface was characterised by slight height variations and numerous trapped gas and/or contamination pockets, which are typical features observed in AFM images of transferred $MoS_2$ flakes[54,55]. When the load force was increased to 120 nN, several regions along the flake perimeter collapsed towards the substrate, reducing the local edge height (Fig. 3a). A further increase in the load force to 150 nN resulted in a pronounced extension of the collapsed areas. Remarkably, once this initial rearrangement occurred, further modifications required only a minimal load force of ~1 nN as evident in the series of images shown in Fig. 3a. This behaviour indicates that the flake was originally suspended on a contamination layer, which as a result of locally applied tip force was pushed away, allowing the $MoS_2$ regions near the edge to establish contact with the Au. After this first point of attachment formed, even a weak stimulus promoted further lamination of the flake onto the substrate.

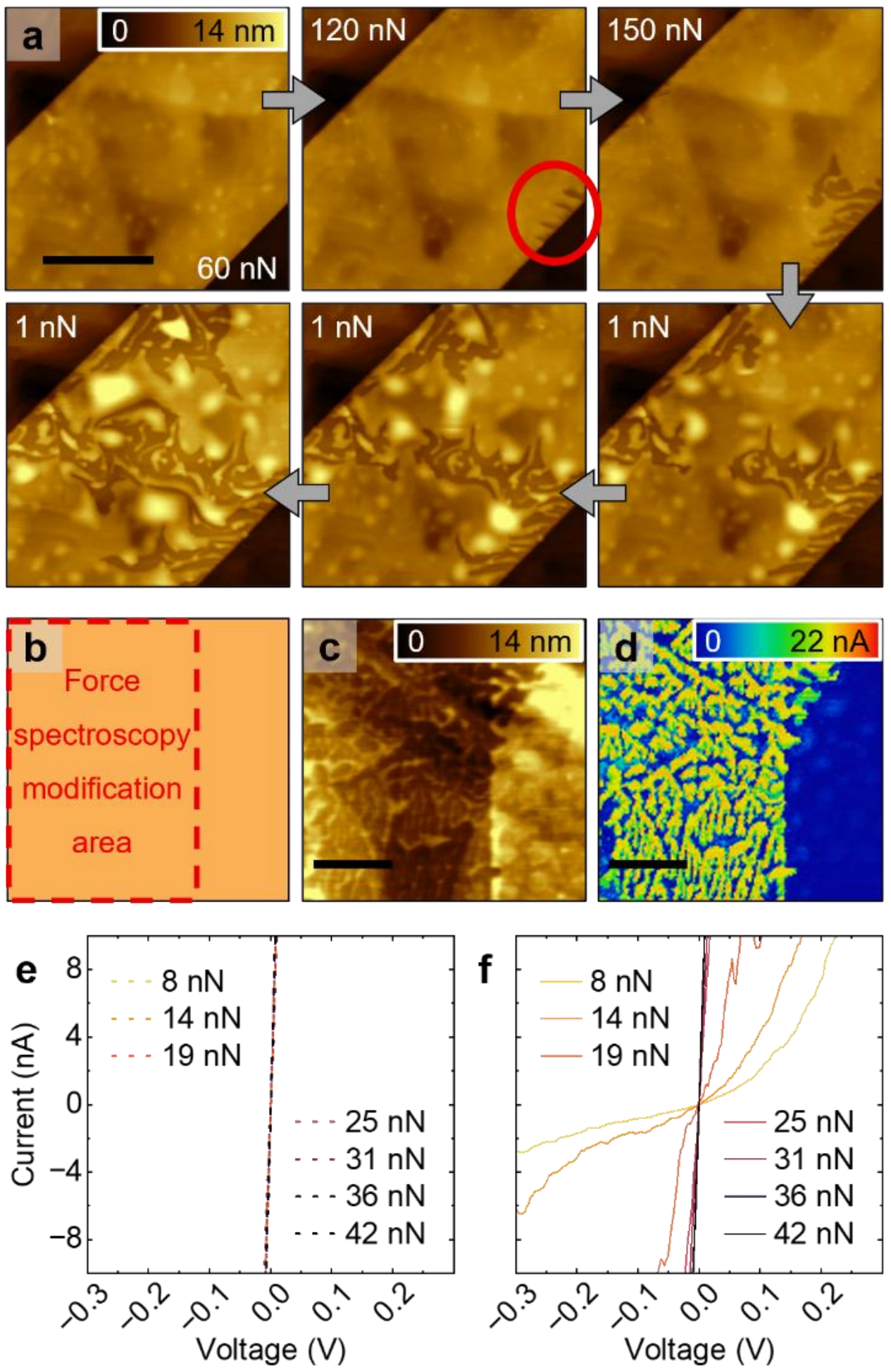

**Figure 3.** Local AFM force spectroscopy $MoS_2$/Au interface modification. (**a**) The process of $MoS_2$ flattening. Scale bar, 1 µm. The load force that was applied by the tip on the entire scan is displayed in the corners. The red circle indicates the area of the start of the flattening process. (**b**) Schematic of the CAFM scan area. (**c**) AFM topography map of the electrical measurement region. Scale bar, 3 µm. (**d**) CAFM current map obtained with bias 0.2 V. Scale bar, 3 µm. (**e**) A set of IV curves measured with a varying load force of the tip in the unmodified area, and (**f**) in the modified area.

Notably, this behaviour was observed only for samples fabricated in an Ar environment (glovebox), where the substrates were initially preheated before deposition to remove residual surface contaminants (see Methods for details). No such effect was observed in ex situ samples. This suggests that heating the substrate in an Ar environment before the $MoS_2$ transfer resulted in the creation of an intermediate interface state: less contaminated than the ex situ samples, yet not as pristine as those prepared under UHV conditions. Such treatment enabled the enhancement of the $MoS_2$/Au contact through the nanoscale lamination technique.

The appearance of bubbles after the modification indicates that some impurities were not expelled but instead relocated, likely increasing strain in the material. This may result from the strong interaction between Au and flattened areas of $MoS_2$, which, once in contact, adhere strongly to each other, preventing the impurities inside the bubbles from returning to their original positions. The high strength of the Au and $MoS_2$ interaction is well documented in the literature and is known to promote adhesion and high-yield exfoliation of TMDs[37,56,57].

The region selected for electrical measurements is shown in Fig. 3b–d. Figure 3b presents a schematic representation of the scan area where the red dashed line marks the region of $MoS_2$ modified by force spectroscopy. Figure 3c shows the AFM topography of the same area, while Fig. 3d displays the corresponding CAFM conductivity map. The conducted CAFM measurement reveals a direct correlation between morphology and electronic behaviour of the layer. The modified region exhibits significantly higher conductivity, with the recorded current being approximately 30 times higher. Representative IV curves measured under varying tip load forces for both the modified and unmodified areas are shown in Fig. 3e and Fig. 3f, respectively. In the first case, the IV characteristics reveal a linear current–voltage relationship with no current dependence on the load force. This is consistent with the UHV-prepared sample, which has a clean interface. In contrast, in Fig. 3f, we observe low conductivity and a nonlinear IV relationship at small tip load forces, in agreement with the ex situ sample, indicating the presence of interfacial contaminants. Upon stronger load force setpoints, the measured current increases linearly, as the impurities are being driven away from the $MoS_2$/Au interface at the probe contact point.

These observations highlight the central role of the interfacial quality in charge transport through ultrathin layers of 2D materials. They indicate that nanoscale variations in interfacial cleanliness dictate whether the junction behaves as a barrier-limited contact or as a pristine ohmic interface. By directly visualising and manipulating the adhesion front at the $MoS_2$/Au boundary, we demonstrate that interfacial contamination - not intrinsic properties of the material - dominates the electronic behaviour commonly reported in the literature. Crucially, the ability to locally laminate $MoS_2$ onto Au and displace trapped impurities shows that the contact quality can be improved *in situ* and with nanoscale precision, effectively replicating the electrical performance of UHV-obtained samples. In addition, this approach opens up new possibilities for nanoscale patterning, with the potential to create locally defined high- and low-resistivity regions.

**Discussion**

Our results demonstrate that the widely reported high contact resistance in monolayer TMD/metal junctions is not an intrinsic limitation of the material, but rather a consequence of uncontrolled interfacial contamination introduced during transfer. By establishing an atomically clean $MoS_2$/Au interface under UHV conditions, we reveal reproducible ohmic behaviour and high conductivity of the $MoS_2$ monolayer, thereby resolving a key limitation that has hindered the implementation of TMD-based electronic devices. The electronic properties of ex situ sample used for comparison provide direct evidence that contamination introduced during sample processing under ambient conditions leads to the formation of high resistance at the monolayer $MoS_2$/Au junction, which is frequently misinterpreted as arising from a high Schottky barrier.

By reconciling previously contradictory reports on the relationship between the conductivity of $MoS_2$ and the number of layers in its structure, our findings establish a unified framework in which interface resistance emerges as the dominant variable. The qualitative resistor-network model further illustrates how modest changes in interfacial resistance can invert the observed thickness-dependent conductivity trend, underscoring the sensitivity of 2D devices to nanoscale interface quality variations. This shows that insufficient control of interfacial cleanliness can lead to qualitatively different interpretations of the electrical behaviour of mono- and multilayer 2D materials.

Furthermore, we demonstrate that a high-quality $MoS_2$/Au interface can be locally achieved even without exfoliation under UHV conditions by combining substrate annealing and layer transfer in an inert atmosphere with subsequent nanoscale lamination. This directly confirms that the presence of interfacial contamination drastically alters the electrical properties of the 2D materials junctions. At the same time, it demonstrates the existence of an efficient method for permanently improving electrical contact at the nanoscale.

As contact resistance remains a defining challenge for the development of high-performance 2D transistors, flexible electronics, and nanoscale optoelectronic systems, deterministic interface engineering becomes a foundational requirement. Our work, therefore, shifts the focus of 2D device optimisation towards comprehensive control of interface quality, a necessary step for reliable 2D electronics.

## Methods

### *Sample preparation*

All samples were exfoliated onto commercially available 300-nm-thick Au(111) films evaporated on mica (Georg Albert PVD). A high-purity (>99.995%) synthetic $MoS_2$ crystal was purchased from HQ Graphene.

In this work, three types of samples with different levels of interface cleanliness were prepared. As a reference, a sample with an atomically clean interface was fabricated entirely under ultra-high vacuum (UHV) conditions, ensuring the highest possible purity of the $MoS_2$/Au interface. For comparison, the other two samples were prepared either under ambient conditions or in an inert atmosphere using the adhesive-tape exfoliation and PDMS transfer method, a procedure widely employed for the preparation of 2D materials[41,42].

For the UHV sample, all preparation steps were carried out in our UHV deposition system (Supplementary Fig. 1). This setup consists of a rotating manipulator equipped with an exfoliation stamp installed in the exfoliation chamber, which is directly connected to the substrate preparation chamber. That configuration enables transfer of the Au substrate between chambers without breaking the vacuum and allows exfoliation of $MoS_2$ directly onto the clean Au surface.

To obtain a clean interface, Au substrates were first cleaned in the preparation chamber by repeated cycles of 2 keV $Ar^+$ ion sputtering and annealing at 663 K. Subsequently, a $MoS_2$ crystal was exfoliated and brought into contact with the clean Au surface under UHV conditions in the exfoliation chamber. As a result, mono- and multilayer $MoS_2$ flakes were deposited directly onto the Au substrate.

An ex situ sample was fabricated under ambient conditions. In this case, the Au substrate was cleaned in the same way as for the UHV samples, but after it was taken out of the vacuum and exposed to ambient air for more than one hour, allowing airborne contaminants to adsorb onto the surface. The $MoS_2$ crystal was then thinned by repeated adhesive-tape exfoliation and transferred onto a PDMS stamp, which was subsequently brought into contact with the Au substrate to deposit thin flakes.

To obtain an intermediate interface condition, an additional set of samples was prepared in an Ar-filled glovebox ($O_2 \approx 0.1$ ppm, $H_2O \leq 5$ ppm). As the substrates had been exposed to ambient air before being put into the glovebox, they were heated in Ar to 373 K to reduce surface contamination that had accumulated during ex situ handling. After cooling to room temperature, the manufacturing process was performed using the same PDMS transfer procedure as for the ex situ sample. This approach produced interfaces of intermediate cleanliness, enabling controlled local modification of their quality using AFM force spectroscopy.

***Sample characterisation***

All AFM measurements were carried out using the Ntegra Aura system (NT-MDT), employing intermittent contact mode imaging, force spectroscopy, as well as contact mode measurements combined with simultaneous current detection through a biased tip–sample junction (CAFM). CAFM measurements for all samples were performed with conductive Pt/Ir-coated AFM tips under low pressure ($\sim 10^{-1}$ mbar) in a dedicated vacuum chamber with a constant tip–sample load force of 1 nN. Before acquiring the data presented in Fig. 1d and Fig. 1h, the samples were annealed at 373 K for 15 minutes, and a series of CAFM cleaning scans was applied to remove surface moisture and contaminants adsorbed during transfer to the AFM. Such contaminants could otherwise lower the measured current by forming a thin insulating layer on top of the exfoliated material and gold or by adhering to the AFM tip and disrupting the electrical connection[58]. In the case of the UHV sample, the transfer to the AFM system required a longer exposure to ambient conditions than for the ex situ sample, which led to their increased adsorption. Consequently, complete removal of contamination from the exposed gold surface for this sample was not possible, resulting in its non-conductive character, as shown in Fig. 1d. However, the cleaning procedure was sufficient to remove contamination from the $MoS_2$-coated areas, enabling reliable conductivity measurements.

***Kirchhoff's-law resistance network simulations***

Resistor network simulations were carried out with the PySpice Python library. In the in-plane direction, each resistor in the $MoS_2$ layer was set to 500 Ω, while those in the Au layer were set to 1 Ω. The $MoS_2$ resistor planes were connected at each node by 10 kΩ resistors, representing the out-of-plane resistance. The bottom $MoS_2$ plane was coupled to the grounded gold plane via interface resistors, set to 10 kΩ for a low interface resistance and 100 kΩ for simulations with a high interface resistance. Each resistor layer consisted of a 128 × 128 grid. Please note that our simulation is a simplified model of a real system. The resistance values have been selected to maintain a certain logic and reproduce experimental conditions; however, they are not intended to reflect the actual values of the resistances of the physical structure.

**Data availability**

The experimental dataset is available from the corresponding author upon reasonable request.

**Code availability**

The code used for the Kirchhoff's-law resistor network simulations is available from the corresponding author on reasonable request.

**Acknowledgements**

This work was supported by the National Science Centre, Poland (grant no. 2020/38/E/ST3/00293).

**Author contributions**

R.D.: Conceptualization, Methodology, Investigation, Formal analysis, Writing – Original Draft. M.L.S.: Validation, Writing – Original Draft. I.L.: Validation, Writing – Review & Editing. M.P.: Software, Writing – Review & Editing. P.D.: Writing – Review & Editing. P.K.: Writing – Review & Editing. W.K.: Writing – Review & Editing. A.N.: Writing – Review & Editing. W.R.: Writing – Review & Editing. P.J.K.: Supervision, Writing – Review & Editing, Resources. M.R.: Conceptualization, Methodology, Supervision, Project administration, Funding acquisition, Writing – Original Draft.

**Competing interests**

The authors declare no competing interests.

Supplementary Information

# Overcoming contact resistance at metal–2D semiconductor interfaces: atomically clean MoS2/Au ohmic junctions

Rafał Dunal[1]*, Maxime Le Ster[2], Iaroslav Lutsyk[1], Michał Piskorski[1], Paweł Dąbrowski[1], Paweł Krukowski[1], Witold Kozłowski[1], Aleksandra Nadolska[1], Wojciech Ryś[1], Paweł J. Kowalczyk[1] & Maciej Rogala[1]*

[1] Department of Solid State Physics, University of Lodz, Lodz, Poland

[2] HUN-REN Centre for Energy Research, Budapest, Hungary

* Corresponding authors. E-mails: rafal.dunal@edu.uni.lodz.pl, maciej.rogala@uni.lodz.pl

## UHV deposition system

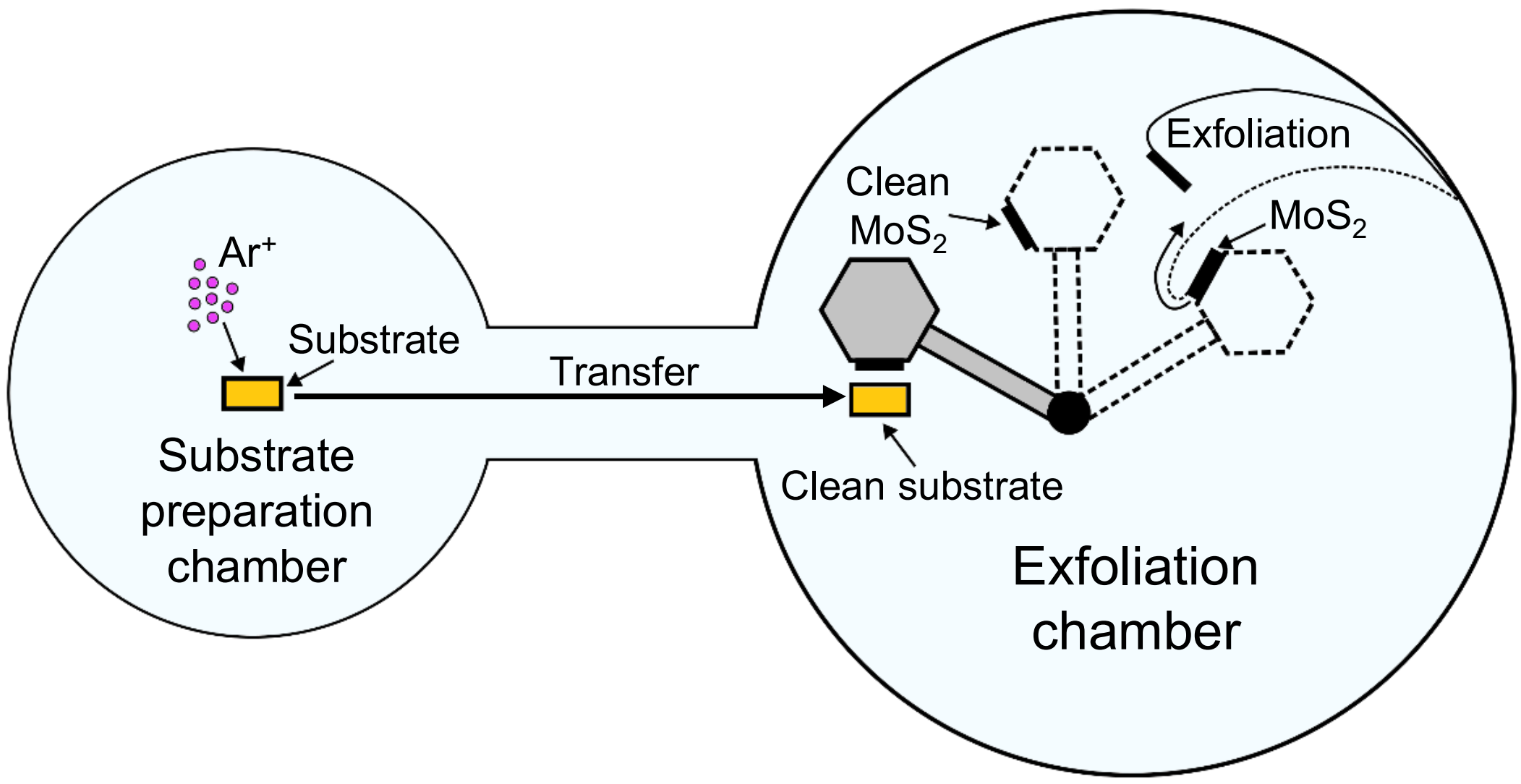

**Supplementary Figure 1.** UHV deposition system. A substrate preparation chamber contains Ar+ sputtering ion gun for sample cleaning. The exfoliation chamber contains a rotating metal arm with a bulk $MoS_2$ crystal mounted with a carbon tape.

Our custom-made exfoliation setup enables Au surface preparation, $MoS_2$ exfoliation, and stamping, all carried out in a continuously maintained UHV environment. This system consists of a preparation chamber for Au substrate cleaning and an exfoliation chamber in which $MoS_2$ is deposited onto a clean Au substrate. Both chambers share the same manipulator that allows the transfer of the samples in UHV conditions. The Au(111) substrate is cleaned by repeated cycles of $Ar^+$ ion sputtering and annealing (two cycles, $Ar^+$ 2 keV, 663K). In the loading/exfoliation chamber, an exfoliation stamp is mounted on a rotating arm as presented. The arm enables the exposure of a clean $MoS_2$ surface through crystal exfoliation and the deposition of thin layers onto clean Au in a single motion.

## $MoS_2$ exfoliation yield in UHV conditions

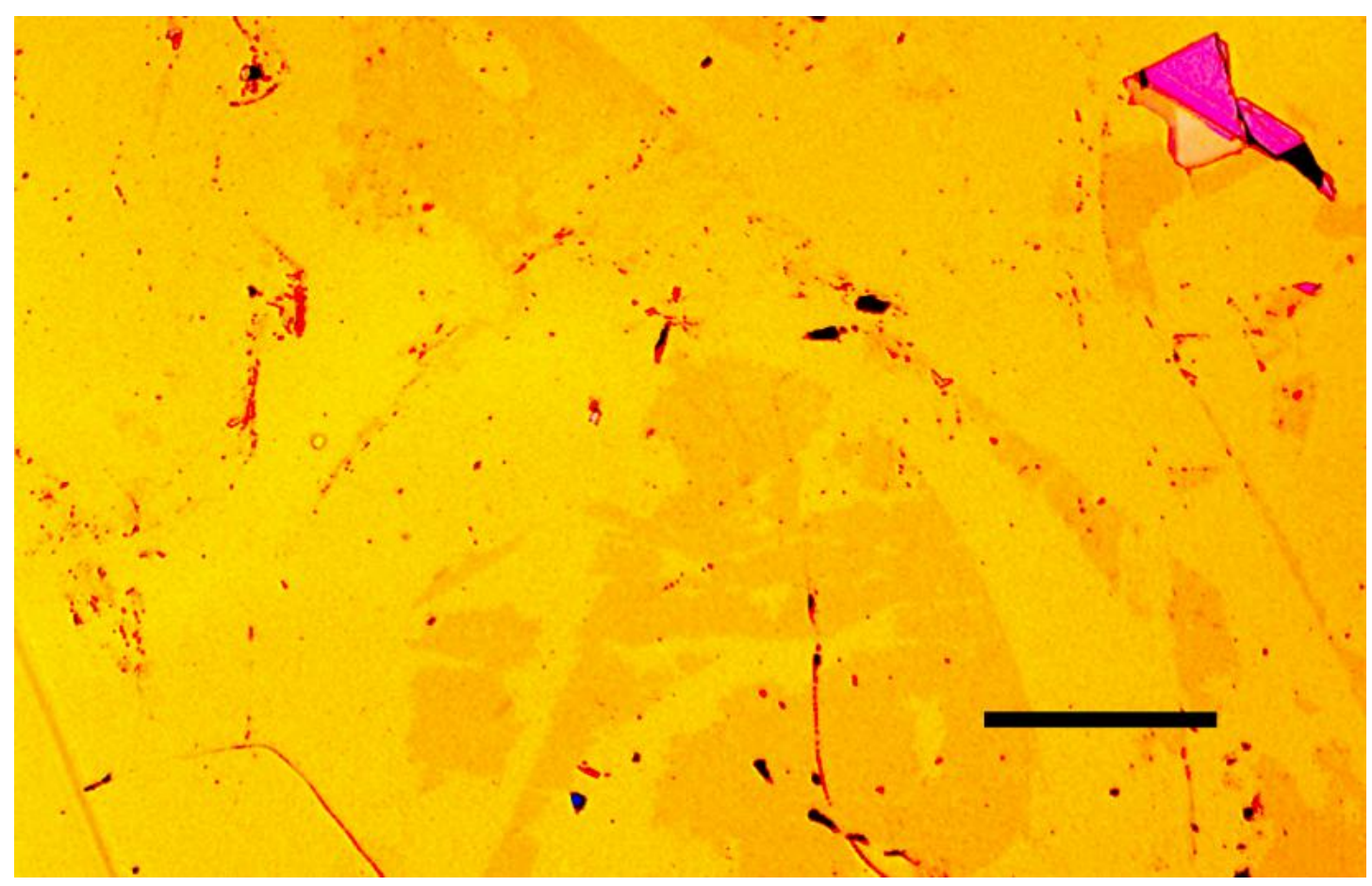

**Supplementary Figure 2.** Optical images of the $MoS_2$ exfoliated onto a gold substrate in UHV conditions (UHV sample). The purity of the interface resulted in a high yield of exfoliation of very thin $MoS_2$ layers. Scale bar, 300 µm.

The optical image depicts the surface of a $MoS_2$/Au sample assembled in UHV, ensuring the formation of a pristine interface. The darker areas correspond to thin $MoS_2$ layers. Their extensive lateral size (>300 µm) indicates excellent adhesion of the $MoS_2$ to Au, which confirms the cleanliness of the interface achieved through Au surface purification using $Ar^+$ ions, followed by the $MoS_2$ exfoliation and deposition in UHV conditions. The few visible scratches result from the manual operation.

**$MoS_2$ UHV sample thickness analysis**

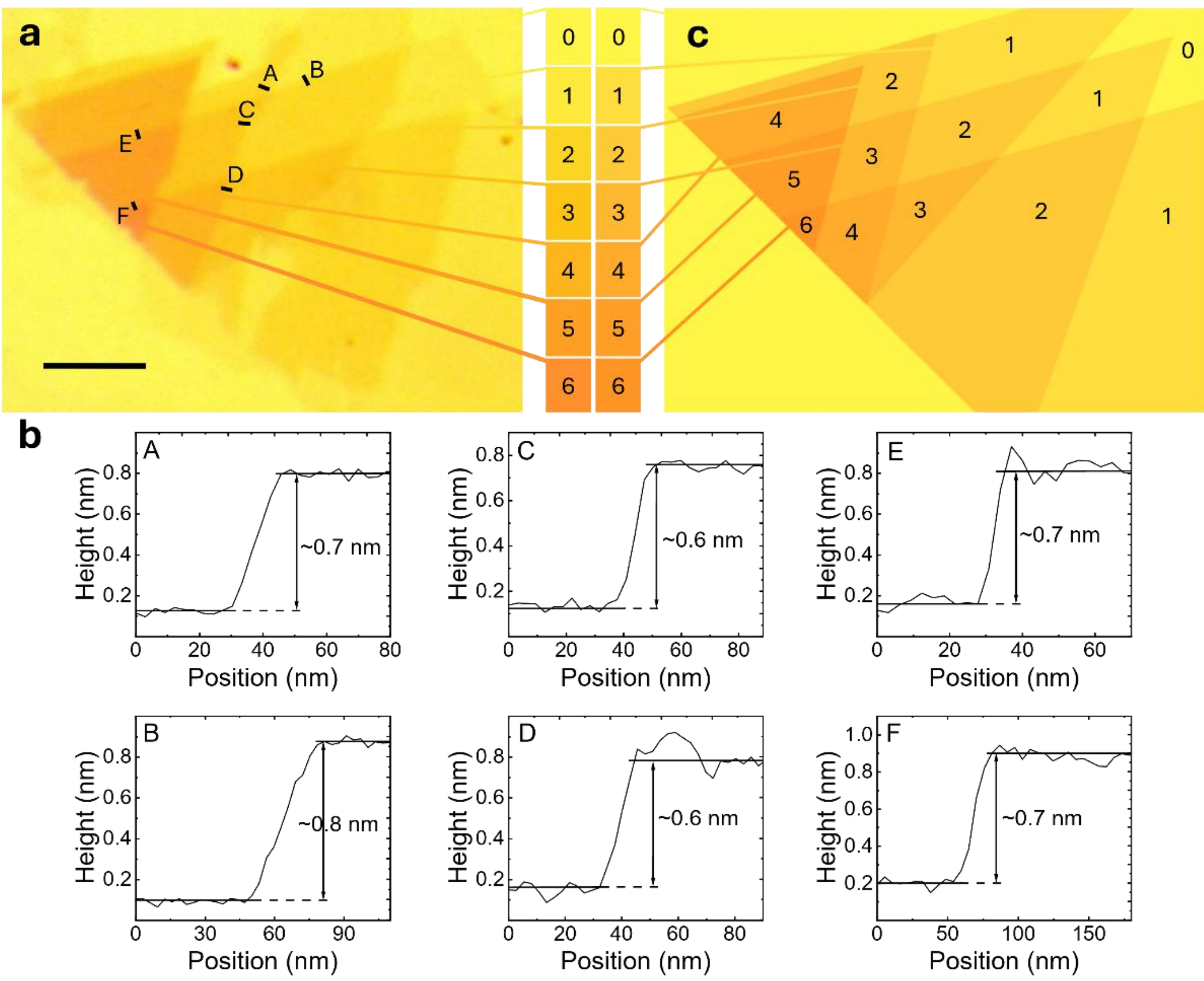

**Supplementary Figure 3.** $MoS_2$ thickness analysis for a UHV sample. (**a**) Optical image of the area selected for CAFM measurements. A colour scale indicates the number of layers for each structure. Scale bar, 5 µm. (**b**) AFM height profiles extracted from locations marked in image (**a**). (**c**) A schematic map of the analysed area with regions labelled according to the assigned number of $MoS_2$ layers.

For the UHV sample, the $MoS_2$ thickness was determined based on AFM topography and optical contrast analysis. The height profiles indicate that each distinct colour corresponds to an increment of one $MoS_2$ layer. Due to contamination of the exposed Au surface during transfer through ambient conditions, AFM height determination of the monolayer with respect to the bare substrate was not possible. Therefore, the layer assignment was based on relative height differences between neighbouring triangular regions and on systematic variations in optical contrast. To visualize the contrast evolution, semi-transparent triangular overlays were digitally applied, enabling direct comparison of colour intensity between adjacent regions. The resulting colour gradation is consistent with the corresponding evolution of optical contrast, confirming the validity of the layer assignment for the individual regions. The consistent optical contrast difference between the regions labelled “1” and “2” in panel c, as well as between the region labelled “1” and the bare Au substrate (labelled “0”), indicates that these correspond to a single-layer increment. Combined with the measured step heights between neighbouring regions (~0.6–0.8 nm), this supports the identification of the region labelled “1” as the monolayer.

**IV characteristics measured for the UHV sample monolayer and reference resistor**

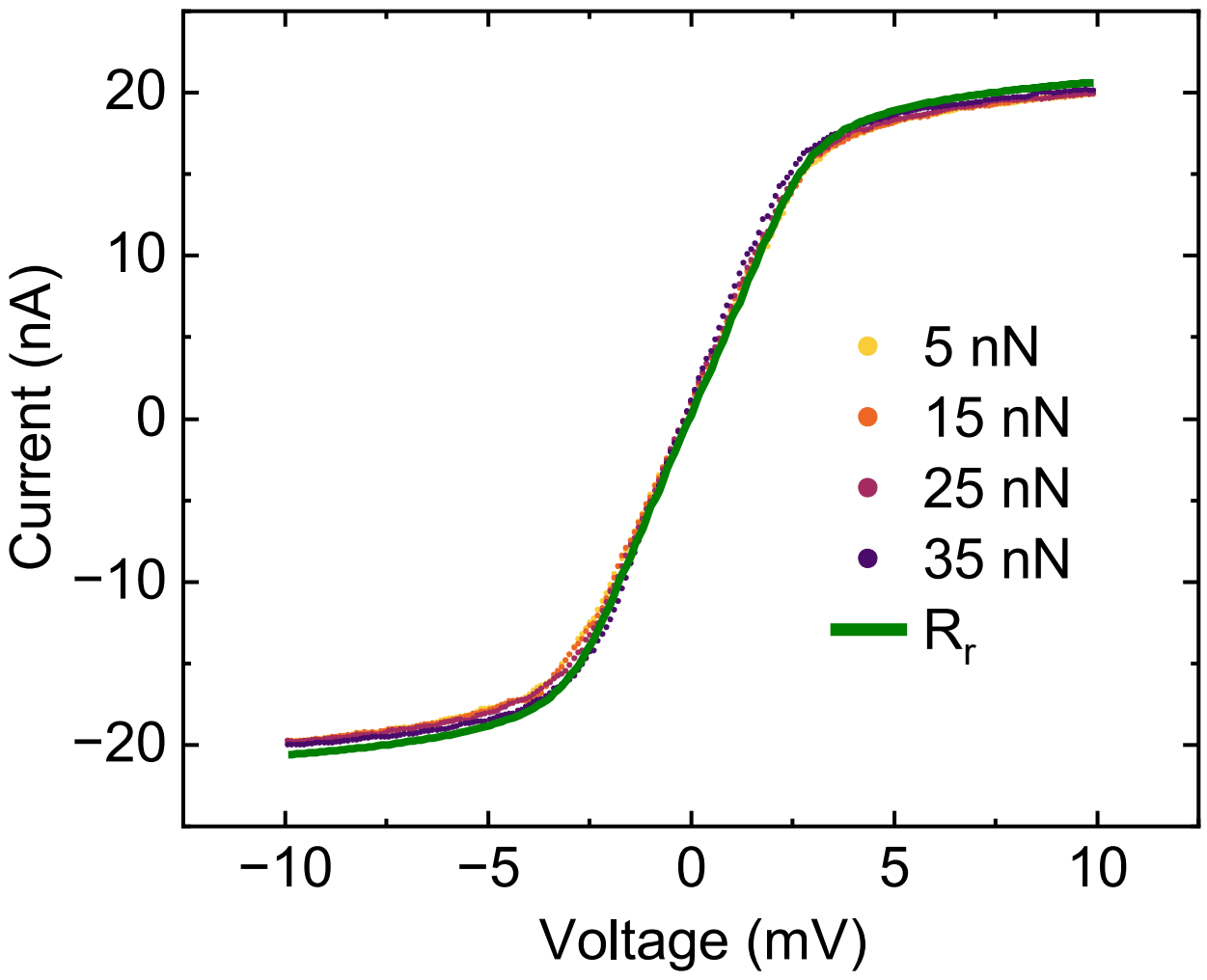

**Supplementary Figure 4.** IV curves measured over the full operating range of CAFM setup for a UHV sample monolayer with varying load force of the CAFM tip and for a 160 kΩ reference resistor.

The electronic architecture of the CAFM microscope used for measurements ensures a strictly linear current readout for tip-sample junction currents below ~10 nA, while at higher current levels the response is nonlinear. This design extends the dynamic range of current detection; however, it prevents a reliable assessment of the linearity of the current–voltage (IV) characteristics over the full applied bias range. To verify the ohmic-like behaviour of the UHV sample $MoS_2$/Au junction, we compared the IV characteristics acquired on the $MoS_2$ monolayer with those obtained for a reference resistor inserted into the measurement circuit in place of the tip–sample contact. Its resistance (~160 kΩ) was chosen to match the effective resistance extracted from the $MoS_2$ I–V data within the linear operating range of the measurement system. The measurements showed that over an extended bias range, the junction response remains nearly indistinguishable from that of the reference resistor, thereby confirming the ohmic-like behaviour of the monolayer $MoS_2$/Au junction.

**IV characteristics measured for the UHV sample**

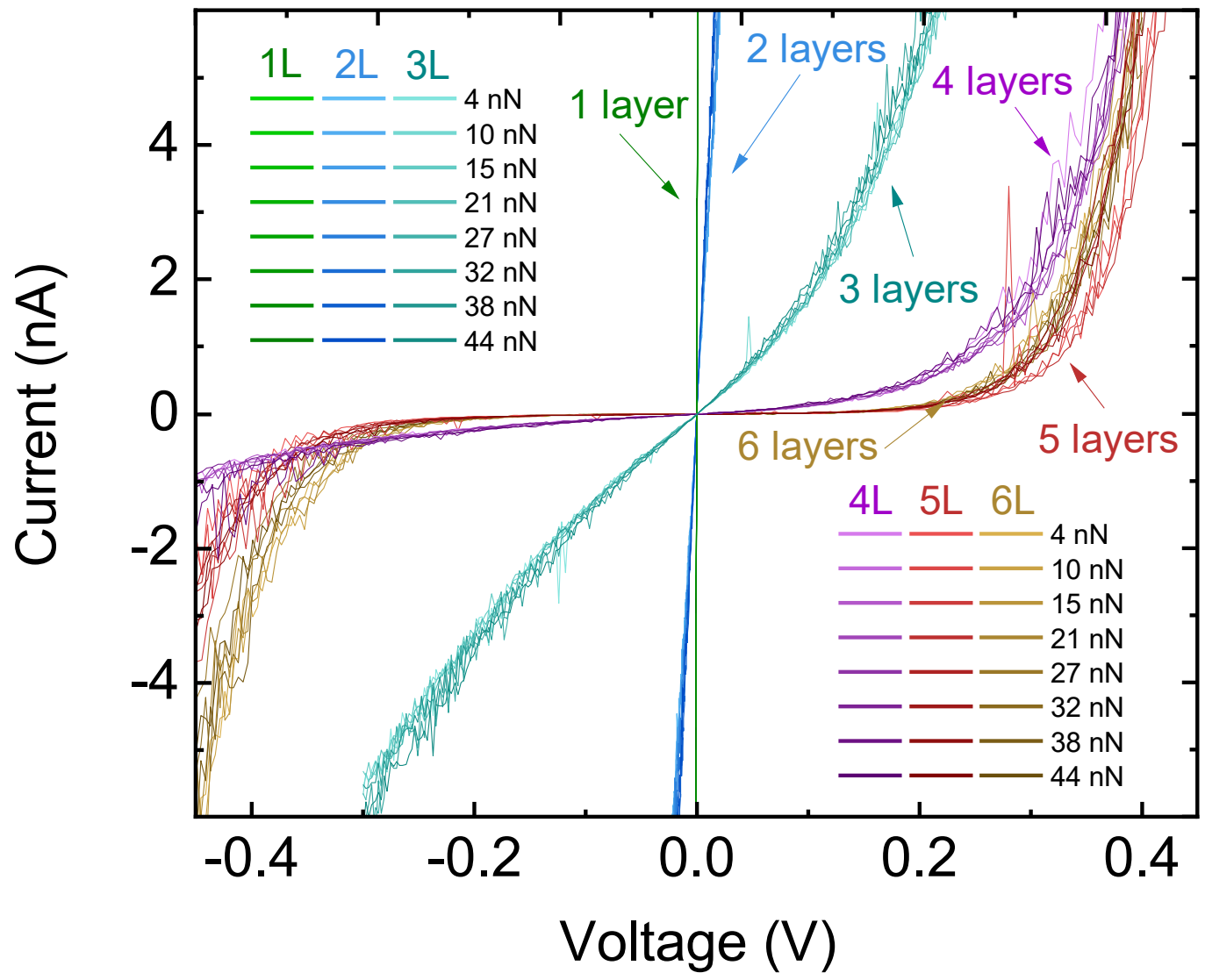

**Supplementary Figure 5.** A representative set of IV curves measured for every $MoS_2$ thickness for a UHV sample with varying load force of the CAFM tip.

The Figure presents a representative set of IV curves measured with the increase of the load force of the tip for every multilayered structure of the UHV sample. The darker the colour of the presented curve, the higher the load force of the CAFM tip. The influence of applied force is negligible, indicating the absence of contaminants at the $MoS_2$-Au substrate interface. Furthermore, the observed high current and ohmic nature of the curves for the first two layers suggest a strong substrate-induced charge transport modulation. Next, for the three layers, the current values are significantly lower, decreasing further for subsequent layers, indicating a decline in conductivity. Additionally, the four-, five-, and six-layer structures exhibit considerable current instabilities during exponential growth, which is particularly noticeable in the case of negative current values.

**Layer-dependent Raman peak shifts of a UHV-prepared $MoS_2$/Au junction**

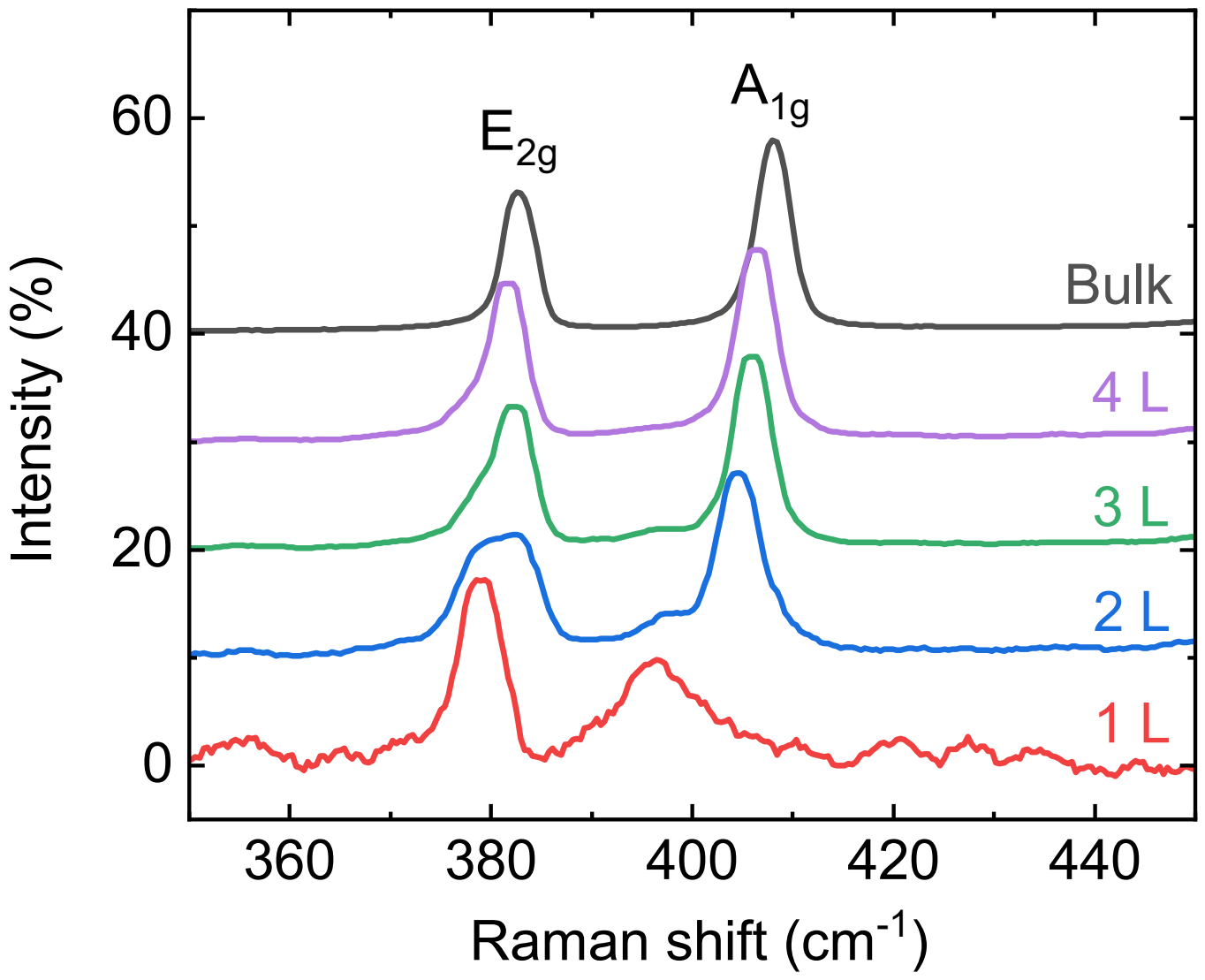

**Supplementary Figure 6.** Raman spectra measured for bulk and 1-4 layer thick $MoS_2$.

For the monolayer, both the $E_{2g}$ and $A_{1g}$ modes exhibit pronounced redshifts and significant broadening compared to the bulk reference, indicating strong interaction with the Au substrate. For 2-3 layer $MoS_2$, both modes recorded for the monolayer regions are still observed, although their intensity decreases. For 4 layers, only a weak remnant of the monolayer $E_{2g}$ mode is observed as a redshift tail of the main $E_{2g}$ mode. These results indicate gradual screening of the substrate with increasing $MoS_2$ thickness. The Raman data are fully consistent with the IV curves measurements present in Supplementary Fig. 4, which reveal strong Au-induced conductivity enhancement in the first three layers, followed by progressive recovery of semiconducting behaviour with increasing thickness.

## Ex situ sample thickness analysis

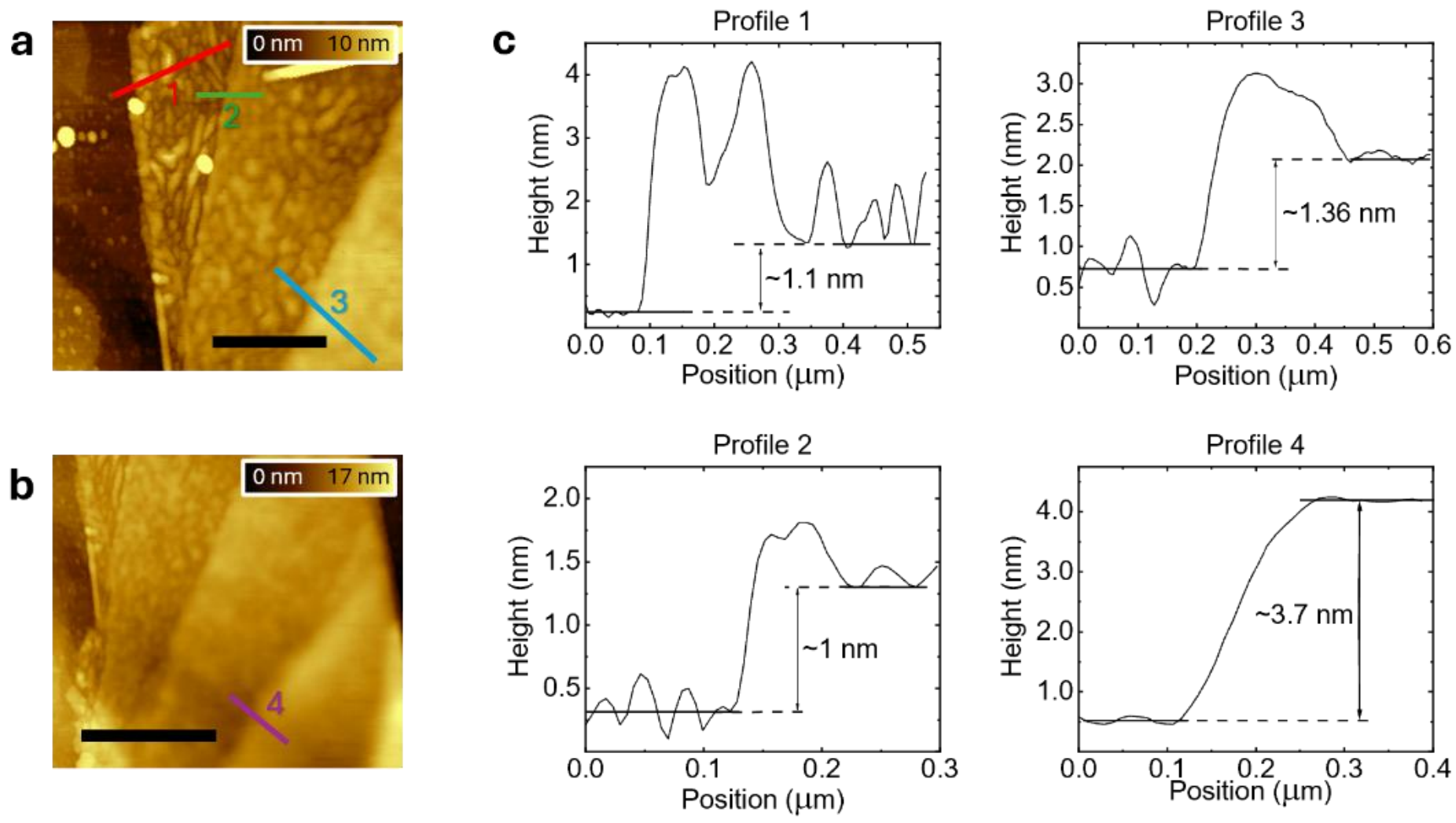

**Supplementary Figure 7.** Ex situ sample thickness analysis. (**a**, **b**) AFM topography scans of the $MoS_2$ layers. Scale bars 500 nm (**a**), 750 nm (**b**). (**c**) Height profile plots extracted from AFM topography scans in (**a**) and (**b**).

The Figure presents height profile plots selected for the estimation of the ex situ sample $MoS_2$ layers' thickness. The high roughness of the thinnest $MoS_2$ layers, with the presence of a large number of bubbles and wrinkles, significantly impedes an accurate measurement, so the exact number of layers can only be estimated. Assuming the literature value of the monolayer thickness of 0.7 nm[1], the value of 1,1 nm for the first layer indicates the presence of the monolayer in the Profile 1 measurement area. The next step is also about 1 nm thick, which implies that the subsequent structure is a double layer. The third layer measured in Profile 3 suggests that the thickness increases by one or at most two layers, but no more. The Profile 4 shows that the next measured thickness is 3,7 nm, which means an increase in the total thickness up to 8 (or 9) layers. It should be noted that the measured profiles are slightly higher than the monolayer values reported in the literature. This increase may result from interfacial contaminants and residues, which contribute to additional measured heights.

**Measured current histograms for $MoS_2$ layers: UHV and ex situ samples.**

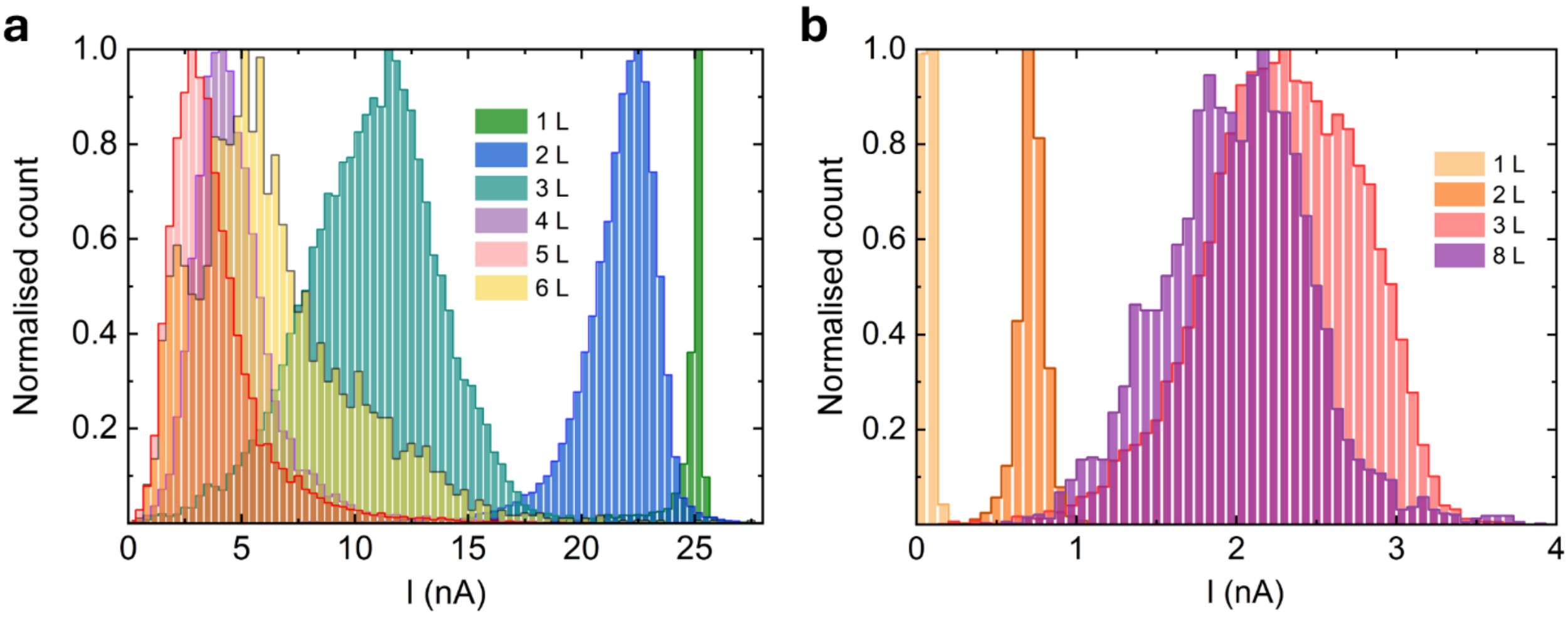

**Supplementary Figure 8.** Normalized current histograms obtained for all measured $MoS_2$ layer thicknesses from CAFM measurements. (**a**) Result for the UHV sample measured at a sample bias of 0.4 V and (**b**) for the ex situ sample at a sample bias of 0.2 V.

The resistance values shown in the main text were determined from the mean currents extracted from the histograms for individual $MoS_2$ layers. The histograms were derived from the current maps shown in Fig. 1d,h of the main text, acquired with a 10 MΩ resistor connected in series. The use of a resistor limited current value and allowed the use of higher bias voltages, enabling simultaneous measurement of both low- and high-resistance regions. For the reported resistance values, the contribution of the 10 MΩ resistor was subtracted.

## Supplementary bibliography